\documentclass[prd,twocolumn,amsmath,amssymb,floatfix,superscriptaddress,nofootinbib]{revtex4-1}

\usepackage{graphicx}
\usepackage{scrextend}
\usepackage{amssymb}
\usepackage{amsmath}
\usepackage{bm}
\usepackage{color}
\usepackage{multirow}
\usepackage{cprotect}

\DeclareMathAlphabet\mathbfcal{OMS}{cmsy}{b}{n}
\definecolor{darkgreen}{RGB}{50,150,0}
\definecolor{purple}{cmyk}{0.5,1.0,0,0}

\def\edth{\;\raise1.0pt\hbox{$'$}\hskip-6pt\partial}
\def\baredth{\;\overline{\raise1.0pt\hbox{$'$}\hskip-6pt
\partial}}

\def\be{\begin{equation}}
\def\ee{\end{equation}}
\def\ben{\begin{equation} \nonumber}
\def\een{\end{equation}}
\def\ban{\begin{eqnarray*}}
\def\ean{\end{eqnarray*}}
\def\ba{\begin{eqnarray}}
\def\ea{\end{eqnarray}}
\def\({\left(}
\def\){\right)}

\newcommand{\bs}{\boldsymbol}

\newcommand{\Msun}{M_\odot}

\definecolor{ultramarine}{rgb}{0.07, 0.04, 0.56}
\definecolor{cadmiumgreen}{rgb}{0.0, 0.42, 0.24}
\definecolor{indigo(dye)}{rgb}{0.0, 0.25, 0.42}
\usepackage[linktocpage=true]{hyperref}
\hypersetup{
colorlinks=true,
citecolor=ultramarine,
linkcolor=cadmiumgreen,
urlcolor=indigo(dye),
pdfauthor={},
pdftitle={},
pdfsubject={}
}

\begin{document}

\title{Observational search for primordial chirality violations using galaxy angular momenta}

\author{Pavel Motloch}
\affiliation{Canadian Institute for Theoretical Astrophysics, University of Toronto, M5S 3H8, ON, Canada}

\author{Ue-Li~Pen}
\affiliation{Canadian Institute for Theoretical Astrophysics, University of Toronto, M5S 3H8, ON, Canada}
\affiliation{Department of Physics, University of Toronto, 60 St. George Street, Toronto, ON M5S 1A7, Canada}
\affiliation{Institute of Astronomy and Astrophysics, Academia Sinica,
Astronomy-Mathematics Building, No. 1, Section 4, Roosevelt Road, Taipei 10617, Taiwan}
\affiliation{Perimeter Institute for Theoretical Physics, Waterloo, N2L 2Y5, ON, Canada}
\affiliation{Canadian Institute for Advanced Research, CIFAR Program in Gravitation and Cosmology, Toronto, M5G 1Z8, ON, Canada}
\affiliation{Dunlap Institute for Astronomy and Astrophysics, University of Toronto, M5S 3H4, ON, Canada}

\author{Hao-Ran~Yu}
\affiliation{Department of Astronomy, Xiamen University, Xiamen, Fujian 361005, China}

\begin{abstract}
\noindent
We search for evidence of primordial chirality violation in the galaxy data
from the Sloan Digital Sky Survey by comparing how strongly directions of galaxy angular
momenta correlate with left and right helical components of a spin vector field
constructed from the initial density perturbations. Within uncertainties, galaxy spins correlate
with these two helical components identically, which is consistent with Universe
without primordial chirality violation. Given current data, it is not yet possible to rule
out maximal chiral violation, although the case of vanishing correlation with the
right helical component is ruled out at about 3.8$\sigma$.
\end{abstract}

\maketitle

\section{Introduction}
\label{sec:intro}

As is well known, interactions of left- and right-handed fermions differ and chiral
symmetry is broken on the microscopic scales (e.g. \cite{Peskin:1995ev}). It is 
then conceivable that chiral symmetry is also broken in the early Universe, for example by
helical couplings between various fields of a multi-field model of inflation. 
Other possible parity breaking mechanisms include addition of the Chern-Simons term to the
gravitational Lagrangian \cite{Lue:1998mq}, gravity at a Lifshitz point
\cite{Takahashi:2009wc} or a chiral gravity with different Newton's constant for the left-
and right-handed gravitational waves \cite{Contaldi:2008yz}. If such violations were
present, they might be manifest at late times and on large scales, for example in the
cosmic microwave background
\cite{Lue:1998mq,Takahashi:2009wc,Contaldi:2008yz,Gluscevic:2010vv}.

In \cite{Yu:2019bsd} we argued that vectors of galaxy angular momenta can also serve as a
useful probe of such a violation. While amplitudes of galaxy momenta are
generally affected by late time evolution and hard to predict (e.g. \cite{Jiang:2018ioo}),
their \emph{directions}\footnote{For brevity, in what follows we use ``spin'' to reference
only the \emph{direction} of the angular momentum of a galaxy / halo.} are rather closely
related to the initial conditions. Indeed, in \cite{Yu:2019bsd} we proposed a vector field
$\bs{J}^\mathrm{IC}$ quadratic in the initial density and gravitational potential (see
Eq.~\eqref{limit_formula} below) that correlates well with spins of dark matter haloes
in numerical simulations and can thus serve as a proxy for halo spins. 
While our simulations focused on dark matter haloes, studies
\cite{2015ApJ...812...29T, Jiang:2018ioo} suggest that spins of galaxies are tracing spins
of their underlying dark matter haloes sufficiently well
to be practically useful as probes of initial conditions.
Currently available galaxy survey data confirms, with statistical significance of about 3$\sigma$, that
galaxy spins indeed correlate with the vector field $\bs{J}^\mathrm{IC}$
\cite{Motloch:2020qvx}. It is thus timely to start searching for the primordial chirality
violations in the galaxy spin data.

With spins of galaxies serving as test probes, in this work we search for signs of 
such violation by comparing correlation strengths of the galaxy spins with the
left and right helical components of $\bs{J}^\mathrm{IC}$, respectively. We define these components below.
In some sense this is analogous to comparing interaction strengths of the left- and
right-handed fermions.

From initial conditions obtained by the ELUCID collaboration in part of the Sloan Digital
Sky Survey (SDSS) volume \cite{2014ApJ...794...94W,Wang:2016qbz}, we calculate the
predicted galactic spin field $\bs{J}^\mathrm{IC}$ and decompose it into its left and
right helical components. We then compare whether the measured galaxy spins correlate with
these two components with identical strength. Nonzero difference would signal either a
parity violation in the early Universe, or a systematic in our measurement.

With sufficient number of galaxy spin measurements, it will be also possible to search for
chirality violations in the galaxy spin field $\bs{J}^g$ directly, without a need to correlate with
$\bs{J}^\mathrm{IC}$ or another similar proxy. For example, one can compare the power
spectra of the left and right helical components of $\bs{J}^g$ \cite{Yu:2019bsd}.
An alternative strategy to search for chirality violation in gallaxy data is comparing
statistical properties of $\bs{J}^\mathrm{IC}_L$ and $\bs{J}^\mathrm{IC}_R$ directly,
without any reference to the galaxy angular momenta (see
Appendix~\ref{sec:appendix} for an example). However, one expects that the optimal way how
to search for these novel effects in the large scale structure data is by
cross-correlating two different observables, as we do here. Unlike searches based on a
single observable, cross-correlations remove the uncorrelated systematics and noise, which
typically leads to favorable detection prospects. This can be seen on numerous historical
examples, such as detection of gravitational lensing of the cosmic microwave background
\cite{Smith:2007rg} and detection of cosmic structure in the 21 cm signal
\cite{2009MNRAS.394L...6P}.

This paper is organized as follows: In \S~\ref{sec:theory} we explain how to construct the
vector field $\bs{J}^\mathrm{IC}$ used to predict the galaxy spins, explain how to
separate it into its left and right helical components and present the correlation
measures and error bar calculation. In \S~\ref{sec:data} we introduce the data used in our
analysis. In \S~\ref{sec:results} we present our results and conclude with a discussion in
\S~\ref{sec:discussion}. Finally, in Appendix~\ref{sec:appendix} we present an
alternative search for a primordial chirality violation that uses only the statistics of
the chiral components of $\bs{J}^\mathrm{IC}$ and does not require any galaxy angular
momenta data.

We denote vectors with bold face letters, their Euclidean norm as $|\bs{v}|$
and label their components with lower case letters $v_i$.
$\epsilon_{ijk}$ and $\delta_{ij}$ represent components of the
three-dimensional Levi-Civita tensor and Kronecker delta, respectively.

\section{Theory}
\label{sec:theory}

In this section we explain how to predict directions of galaxy angular
momenta from initial conditions, introduce the statistics we use to study chiral
symmetry breaking and explain how we quantify uncertainties of our measurements.

\subsection{
Predicting galaxy spins
}

In our current understanding, the dark matter haloes acquire angular momenta from the
inhomogeneous tidal field that torques the non-spherical protohalo early on
\cite{Peebles:1969jm,1970Ap......6..320D,1984ApJ...286...38W,
Porciani:2001db,Porciani:2001er}.
At late times, interactions
with the nearby large scale structure notably complicate the picture
\cite{Porciani:2001db,Porciani:2001er, Krolewski:2019bfv,
2017ApJ...841...16D,Zhang:2014rju,Veena:2018ooo,Veena:2019ozd,Kraljic:2021oeg,Kraljic:2019lca,
Wang:2017tsr,Neyrinck:2019uvc,Schaefer:2008xd,Codis:2012ep,Codis:2015tla,Welker:2019puz,
Jones:2010cs,AragonCalvo:2006ay,Hahn:2007ui,Bett:2011rs,Bett:2015aoa,Tempel:2013gqa,Wang:2018rlf}.

In \cite{Yu:2019bsd} we showed that the vector field $\bs{J}^\mathrm{IC}$ constructed from the
initial density field $\rho$ and the initial gravitational potential field $\phi$
according to
\be
\label{limit_formula}
J^\mathrm{IC}_a
=
\sum_{bck}
\epsilon_{abc}
\partial_{bk} \phi^r
\partial_{kc} \rho^r
\ee
can be used to predict the \emph{directions} of final angular momenta of dark matter halos
in simulations. To suppress fluctuations on scales too small to affect the halo angular
momentum, the initial fields $\rho, \phi$ are smoothed with a Gaussian kernel with a
suitably chosen smoothing scale $r$. This is indicated
by the superscript $r$. With perfectly known initial conditions, the optimal smoothing
scale for a particular halo depends on its mass and for a $10^{13}M_\odot$ halo
corresponds to about $2\, h^{-1}\ \mathrm{Mpc}$.
For each halo, the vector field $\bs{J}^\mathrm{IC}$ should be evaluated at its
Lagrangian centre of mass.

\subsection{Quantifying the chiral violation}

In practice, measurements of halo angular momenta are unavailable and we have to work with
angular momenta of galaxies $\bs{J}^g$.
Given $\bs{J}^g$ and values of $\bs{J}^\mathrm{IC}$ at 
galaxy Lagrangian positions, we can define the correlation strength
\be
\label{define_mu}
  \mu = \left\langle 
  \frac{\bs{J}^g}{|\bs{J}^g|}
  \cdot
  \frac{\bs{J}^\mathrm{IC}}{|\bs{J}^\mathrm{IC}|}
  \right\rangle ,
\ee
with the average taken over the galaxies. In case of no relationship between the
directions of $\bs{J}^g$ and $\bs{J}^\mathrm{IC}$, the correlation between these
two sets of vectors vanishes. 

As explained in greater detail in \cite{Yu:2019bsd}, it is straightforward to
split the divergence-free vector field $\bs{J}^\mathrm{IC}$ into its two helical components
$\bs{J}^\mathrm{IC}_L, \bs{J}^\mathrm{IC}_R$. This decomposition can be
achieved in the Fourier domain, where the components of the transformed field $\tilde J^\mathrm{IC}_a$
are combined using two projection operators $\mathbb{P}^{L/R}$,
\be
  \tilde J^\mathrm{IC}_{L/R,a}(\mathbf{k}) \equiv \sum_b
  \mathbb{P}^{L/R}_{ab}(\mathbf{k}) 
  \tilde J^\mathrm{IC}_b(\mathbf{k}) .
\ee
These projection operators explicitly read
\be
  \mathbb{P}^{L/R}_{ab}(\mathbf{k}) = 
  \frac{1}{2}
  \left[
  \(\delta_{ab} - \hat k_a \hat k_b\) 
  \pm 
  i \sum_c \epsilon_{abc} \hat k_c
  \right]
\ee
and we defined the unit vector $\bs{\hat k} = \bs{k}/|\bs{k}|$. The real space components
of $\bs{J}^\mathrm{IC}_{L/R}$ are then obtained through a backward Fourier transform.
Under the parity transformation the two helical components swap,
\ba
\label{parity_swap}
  \bs{J}^\mathrm{IC}_L \leftrightarrow \bs{J}^\mathrm{IC}_R .
\ea

Analogously to $\mu$, we can define the correlations with the left- and right-handed
helical components of $\bs{J}^\mathrm{IC}$ as
\be
\label{define_mu_LR}
  \mu_X = \left\langle 
  \frac{\bs{J}^g}{|\bs{J}^g|}
  \cdot
  \frac{\bs{J}^\mathrm{IC}_X}{|\bs{J}^\mathrm{IC}_X|}
  \right\rangle \ \ \ \ \ X \in \{L, R\}
\ee
and their difference,
\be
  \mu_- = \mu_L - \mu_R .
\ee
Because of the swap of $\bs{J}^\mathrm{IC}_L$ and $\bs{J}^\mathrm{IC}_R$ under 
parity transformation, $\mu_-$ flips sign. In a universe where
the chiral symmetry is broken at high redshifts and this violation propagates into galaxy
spins, we would detect nonzero $\mu_-$. On the other hand, in standard model we
expect to find $\mu_-$ consistent with zero. 

At present, it is possible to obtain the full direction vector of angular momentum 
$\bs{J}^g$ for only a limited number of galaxies as it is necessary to perform an integral
field spectroscopy and provide additional information. Measurements of $\mu_-$
would thus be too noisy.
Fortunately, for our purposes of searching for the primordial chirality violation, we do
not need to know the full direction vector $\bs{J}^g/|\bs{J}^g|$. By focusing only on the
components of $\bs{J}^g$ along the line of sight direction $\bs{\hat r}$, we can significantly
increase the number of galaxies in our sample.

When we restrict our attention to only the line of sight component of the galaxy spin,
\be
  \bs{J}^g \rightarrow 
  \left(\bs{J}^g \cdot \bs{\hat r}\right)
  \bs{\hat r} 
  ,
\ee
the vector $\bs{J}^g/|\bs{J}^g|$ in Eq.~\eqref{define_mu} and \eqref{define_mu_LR} is
replaced with
\be
\label{spin_sign}
  \frac{\bs{J}^g}{|\bs{J}^g|} \rightarrow 
  \bs{\hat r}\,
  \mathrm{sign}( \bs{J}^g \cdot \bs{\hat r})
  = \pm \bs{\hat r}
  .
\ee
Rotational state of each galaxy is thus represented by a single bit of information,
instead of a unit vector.  For a later notational convenience we introduce a shorthand
\be
  S \equiv \mathrm{sign}( \bs{J}^g \cdot \bs{\hat r}) .
\ee

Analogously to above, we can define correlation strength with the reconstructed spin field
\be
\label{define_mu_par}
  \mu^\parallel = \left\langle 
  S \bs{\hat r}
  \cdot
  \frac{\bs{J}^\mathrm{IC}}{|\bs{J}^\mathrm{IC}|}
  \right\rangle .
\ee
Using SDSS data, we experimentally verified that
$\bs{J}^\mathrm{IC}$ defined in Eq.~\eqref{limit_formula} leads to nonzero $\mu^\parallel$ with a
statistical significance of about $3\sigma$ \cite{Motloch:2020qvx}.

For the purposes of searching for the primordial chirality violation we also define
correlations with the chiral components of $\bs{J}^\mathrm{IC}$
\be
\label{define_mu_LR_par}
  \mu_X^\parallel = \left\langle 
  S \bs{\hat r}
  \cdot
  \frac{\bs{J}^\mathrm{IC}_X}{|\bs{J}^\mathrm{IC}_X|}
  \right\rangle \ \ \ \ \ X \in \{L, R\}
\ee
and finally also
\be
  \mu_-^\parallel = \mu_L^\parallel - \mu_R^\parallel .
\ee

Notice that even after the restriction to radial components, $\mu_-^\parallel$ remains sensitive to
chirality violations. Measurement of $\mu_-^\parallel$ thus amounts to indirectly probing
chiral symmetry in the early Universe and will be the main result of this work.

\subsection{Error bars}

Our measurement of $\mu_-^\parallel$  are based on the vector fields
$\bs{J}^\mathrm{IC},\bs{J}^\mathrm{IC}_{L},\bs{J}^\mathrm{IC}_{R}$ built up from the
reconstructed initial
conditions and a set of measured galaxy positions
$\{\bs{r}_1, \bs{r}_2, \dots \bs{r}_N\}$ and signs of the radial components
of their angular momenta $\{S_1, S_2, \dots S_N\}$. Here $N$ is the number of galaxies in
our sample.

To estimate the uncertainties of our measurements, we repeat the calculations with the
same vector fields $\bs{J}^\mathrm{IC},\bs{J}^\mathrm{IC}_{L,R}$ and galaxy positions
$\{\bs{r}_1, \bs{r}_2, \dots \bs{r}_N\}$ but randomly shuffled signs of the radial
component of the galaxy angular momenta,
\be
\{S_{1}, S_{2}, \dots S_{N}\} \rightarrow
\{S_{\sigma_1}, S_{\sigma_2}, \dots S_{\sigma_N}\} ,
\ee
where $\sigma_i$ is some permutation of $\{1, 2, \dots N\}$. We repeat the
calculations 40 000 times with independent random $\sigma_i$ and as uncertainties of our
results then take the standard deviations of these randomized results.

We estimate the error bars similarly for $\mu^\parallel, \mu_L^\parallel$ and
$\mu_R^\parallel$.

\section{Data}
\label{sec:data}

In this section we present the observational data used in this work. We start by
describing the initial density field $\rho$ as reconstructed by the ELUCID collaboration.
We then introduce the data used to determine angular momenta of
galaxies and their positions.

\subsection{Initial conditions}

The initial density field $\rho$ used in this work was obtained by the ELUCID
collaboration \cite{2014ApJ...794...94W,Wang:2016qbz}.

They first pre-processed SDSS data to create a catalog of galaxy
groups \cite{Yang:2007yr} and then determined mass of each group via a luminosity-based
abundance matching. They corrected for peculiar velocities and only retained groups in the
Northern Galactic Cap, redshift range $0.01 \le z \le 0.12$ and with masses above
$10^{12}\Msun$. The space was then tessellated
according to which galaxy group was the closest. Within the resulting sub-volumes,
particles were
placed randomly, in accordance with the expected density profile for halo of given mass.
This particle distribution represents today's density field.

In the second step of the reconstruction, ELUCID collaboration ran a Particle-Mesh (PM)
dynamics code repeatedly in a Hamiltonian Monte Carlo fashion to determine the
best fit initial conditions. For each random set of initial conditions, the PM code was
used to calculate the corresponding value of today's density field. By comparing this
density field with that determined from the SDSS data, it was possible to construct a
probability measure on the space of the initial conditions. Due to inaccuracies of the PM
code on small scales, both density fields were smoothed on a scale of $4\, \mathrm{Mpc/h}$
before comparison. Iteratively probing the space of initial conditions then allowed ELUCID
to find the initial conditions that best describe the local galaxy data.

From these best fit initial conditions, we calculate the initial gravitational
potential $\phi$ from the Poisson equation and use \eqref{limit_formula} to
predict the galaxy spins $\bs{J}^\mathrm{IC}$.

\subsection{Galaxy Spins}

We base our determination of the sign of $\bs{J^g} \cdot \bs{\hat r}$ on the fact that for
spiral galaxies the orientation of the angular momentum of the galaxy's gas is closely
related to the sense of rotation of galaxy's spiral arms (clockwise or anti-clockwise, i.e.
in the sense of the letters Z or S).
To determine the direction of the radial component of $\bs{J}^g$, it is thus sufficient to determine
whether given galaxy rotates clockwise or anti-clockwise, with the radial component of
galaxy spin aligned ($S = 1$) resp. anti-aligned ($S = -1$) with its position vector.
This classification is not a perfect determination of the galaxy angular momentum, with
about 4\% of galaxies having angular momentum that is pointed in the opposite direction
than that inferred from the orientation of the spiral arms \cite{1982Ap&SS..86..215P}, but
this effect is not expected to bias the results.

We use a catalog of galaxies classified as ``clockwise spiral galaxy'' and ``anticlockwise
spiral galaxy'' by Galaxy Zoo \cite{Lintott:2008ne}, a citizen science project
where members of the public visually classified properties of almost $9 \times 10^5$
objects. For each object, summary statistics of the voting results are
publicly available and we obtained them through
CasJobs\footnote{https://skyserver.sdss.org/CasJobs/}. We only consider objects classified
by at least 80\% of votes as either ``clockwise spiral galaxy'' or
``anti-clockwise spiral galaxy''. 
Our final catalog contains 12022 galaxies and the corresponding values of $S$.

\subsection{Galaxy Positions} 

From CasJobs we also obtain redshift and sky position for each galaxy in our sample.
This allows us to find the galaxy's three-dimensional position and use it to interpolate
$\bs{J}^\mathrm{IC}$, making use of an inverse displacement field of the
reconstructed simulation. The line of sight vector $\bs{\hat r}$ is also
obtained directly from the galaxy position.

\section{Results}
\label{sec:results}

To construct the vector field $\bs{J}^\mathrm{IC}$, we need to choose a scale $r$ with
which to smooth the initial conditions. For this work, we choose $r = 3\,
h^{-1}\mathrm{Mpc}$, as it is the smoothing scale that for current data leads to the
maximal correlation $\mu^\parallel$ in \cite{Motloch:2020qvx}.

Using the (anti-)clockwise classifications of the SDSS galaxies and the initial
conditions as determined by the ELUCID collaboration, we find
\ba
  \mu_L^\parallel &=& \(0.41 \pm 0.53\) \times 10^{-2}\\
  \mu_R^\parallel &=& \(1.99 \pm 0.53\) \times 10^{-2} .
\ea
The parity-odd variable $\mu_-$ is then
\be
  \mu_-^\parallel = \(-1.58 \pm 0.75\) \times 10^{-2},
\ee
formally a 2.1 $\sigma$ deviation from the value of zero expected a in parity invariant
universe.
Notice that the error bars suggest that $\bs{J}^\mathrm{IC}_L/|\bs{J}^\mathrm{IC}_L|$
and $\bs{J}^\mathrm{IC}_R/|\bs{J}^\mathrm{IC}_R|$ are essentially uncorrelated, which we
also checked explicitly.

The parity even combination
\be
  \mu_+^\parallel \equiv \mu_L^\parallel + \mu_R^\parallel = \(2.40 \pm 0.74\) \times 10^{-2} .
\ee
is detected with similar significance as $\mu^\parallel$,
\be
  \mu^\parallel = \(1.80 \pm 0.53\) \times 10^{-2} .
\ee
The two are not identical despite
\be
\label{JIC_sum_split}
  \bs{J}^\mathrm{IC} = 
  \bs{J}^\mathrm{IC}_L +
  \bs{J}^\mathrm{IC}_R ,
\ee
as normalization of the vectors in Eq.~\eqref{define_mu_par} and \eqref{define_mu_LR_par} is a
nonlinear operation.

\section{Discussion}
\label{sec:discussion}

In this work we searched for signature of parity violations in the angular momenta
vectors of the SDSS galaxies. We found a mild preference for galaxy spins to correlate more
strongly with the right helical component of the vector field $\bs{J}^\mathrm{IC}$
built from the initial density field according to \eqref{limit_formula}, but this
preference is not statistically significant for the currently available galaxy sample and
initial condition reconstruction. Given current uncertainties, the data is thus
consistent with no parity violation. At the same time, our result is at present also
consistent with the maximally violating case $\mu_L^\parallel = 0, \mu_R^\parallel \neq 0$, while the other
maximally violating case $\mu_R^\parallel = 0, \mu_L^\parallel \neq 0$ is excluded at about $3.8\sigma$.

In principle, there are other parity-odd observables one can construct beyond
$\mu_-^\parallel$.
Another possibility would be for example 
\be
\label{alternative_formula}
  \Delta \mu^\parallel = \left\langle 
  S \bs{\hat r}
  \cdot
  \left(
  \frac{\bs{J}^\mathrm{IC}_L - \bs{J}^\mathrm{IC}_R}{|\bs{J}^\mathrm{IC}|}
  \right)
  \right\rangle ,
\ee
which differs from $\mu^\parallel$ by only a sign in the numerator (see
\eqref{JIC_sum_split}).
However, this statistic is quite sensitive to galaxies with small
$|\bs{J}^\mathrm{IC}|$, which leads to a strongly non-Gaussian distribution and long tails.
The reader can
contrast this with $\mu_{L/R}^\parallel$ that are both limited to $[-1, 1]$ and
no single galaxy can dominate the $\mu_-^\parallel$ statistics, which is the reason behind our
choice.

We need to stress that the error bar estimates quoted in this work represent only the
statistical uncertainty. In principle, there can also be systematic uncertainties
affecting our results. These can arise for example from biases of the human observers
classifying the galaxies as clockwise/anticlockwise \cite{2017MNRAS.466.3928H} or 
from the scanning strategy of the survey, that can break the parity in subtle
ways. Given that in this work we found no statistically significant chirality violation,
we do not attempt to perform the involved computations that would be necessary to estimate
these systematic uncertainties.

In the nearest future, data from Dark Energy Spectroscopic Instrument \cite{Levi:2013gra}
will allow us to notably shrink the error bars of the $\mu_-^\parallel$ measurement.  Improving the
reconstruction on smaller spatial scales will be especially interesting, because
simulations suggest that the quadratic formula~\eqref{limit_formula} can lead to
correlations up to an order of magnitude stronger than what is currently achievable.
Additional improvements are then expected from leveraging the extensive theoretical and
observational knowledge of the origin of galaxy spins 
\cite{Porciani:2001db,Porciani:2001er,Schaefer:2008xd,Codis:2012ep, AragonCalvo:2006ay,Hahn:2007ui,Codis:2012ep,Bett:2011rs, Bett:2015aoa,Tempel:2013gqa,Veena:2018ooo,Veena:2019ozd, Wang:2018rlf,Wang:2021axr,GaneshaiahVeena:2020bsx,Xia:2020znt, Jones:2010cs,Krolewski:2019bfv,Welker:2019puz} 
to further tighten the relationship
between the galaxy spins and initial conditions and improve on the simple formula
Eq.~\eqref{limit_formula}.
Another interesting follow-up work would be to pick a particular model of chiral breaking and
propagate this breaking all the way to the final galaxy spins, to get an estimate
of how big an effect one might expect to observe.

While this draft was being finalized, a related work \cite{Cahn:2021ltp} appeared, where
an alternative strategy to search for primordial chirality violations (using
four-point functions of the galaxy density field) was introduced.

\acknowledgements{
We thank an anonymous referee for numerous useful suggestions including motivating the
Appendix and J.~Braden for useful
discussions. 

This work was supported by Natural Sciences and Engineering Research
Council of Canada (NSERC) grant CITA 490888. 
P.M. was additionally supported by Vincent and Beatrice Tremaine Fellowship.
U.-L.P. received
additional support from Ontario Research Fund-research Excellence Program grant RE09-024,
NSERC grants RGPIN-2019-067, 523638-201, CRDPJ 523638-2, Canadian Institute for Advanced
Research grants FS21-146 and APPT, Canadian Foundation for Innovation grant IOF-33526,
Simons Foundation grant 568354, Thoth Technology Inc., and the Alexander von Humboldt
Foundation.
H.-R.Y. additionally acknowledges support
from National Natural Science Foundation of China grant 11903021. 
}

\bibliography{ic_primordial_chirality}
\appendix
\section{Alternative statistic}
\label{sec:appendix}

In principle, any potential chiral violation present in our Universe can also cause
differences between the statistical properties of the reconstructed vector fields
$\bs{J}^\mathrm{IC}_L$ and $\bs{J}^\mathrm{IC}_R$. For example, it is possible
to compare variances of these vector fields by evaluating
\be
  \eta = \frac{
    \left\langle \left|\bs{J}^\mathrm{IC}_L\right|^2\right\rangle_V
    -
    \left\langle \left|\bs{J}^\mathrm{IC}_R\right|^2\right\rangle_V
    }
    {
    \left\langle \left|\bs{J}^\mathrm{IC}\right|^2\right\rangle_V
    } ,
\ee
where the subscript reminds us we are averaging only over the volume in which we have
reconstructed initial conditions and we normalize by the variance of $\bs{J}^\mathrm{IC}$
for convenience. Detecting nonzero $\eta$ would suggest either an uncorrected systematic,
or a sign of primordial chirality violation.

With vector fields $\bs{J}^\mathrm{IC}, \bs{J}^\mathrm{IC}_L, \bs{J}^\mathrm{IC}_R$
obtained from the reconstructed initial conditions smoothed with the smoothing scale $r =
3\, h^{-1}\ \mathrm{Mpc}$, we get
\be
\label{eta_result}
\eta = (-0.8 \pm 1.4)\times 10^{-3} .
\ee
To estimate the error bar, we ran 14 simulated $\Lambda$CDM universes, calculated $\eta$
in each and took the standard deviation of these results. These simulations were
performed using the $N$-body code CUBE \cite{Yu:2017rfx} and were run on a $512^3$ grid
representing volume $(500 h^{-1}\mathrm{Mpc})^3$.
Unlike in the data, we do not perform any reconstruction and build $\bs{J}^\mathrm{IC}$
from the true initial conditions. This means the error bars in \eqref{eta_result} are
somewhat underestimated.

A more rigorous analysis, including comparing power spectra of $\left|
\bs{J}^\mathrm{IC}_L \right|$ and $\left|\bs{J}^\mathrm{IC}_R \right|$, would necessarily
involve deconvolving the window function and goes beyond the scope of the present work.

\end{document}